\documentclass{article}

\PassOptionsToPackage{numbers, compress}{natbib}
\usepackage[final]{neurips_ts4h_2022}

\usepackage[utf8]{inputenc} % allow utf-8 input
\usepackage[T1]{fontenc}    % use 8-bit T1 fonts
\usepackage{hyperref}       % hyperlinks
\usepackage{url}            % simple URL typesetting
\usepackage{booktabs}       % professional-quality tables
\usepackage{amsfonts}       % blackboard math symbols
\usepackage{nicefrac}       % compact symbols for 1/2, etc.
\usepackage{microtype}      % microtypography
\usepackage{xcolor}         % colors
\usepackage{graphicx}

\title{Improving ECG-based COVID-19 diagnosis and mortality predictions using pre-pandemic medical records at population-scale}

\author{
Weijie Sun$^{1,2}$ \And Sunil Vasu Kalmady$^{1,2,3}$ \And Nariman Sepehrvand$^{1,4}$ \And Luan Manh Chu$^{1,4}$ \And Zihan Wang$^5$ \And Amir Salimi$^2$ \And Abram Hindle$^2$ \And Russell Greiner$^{2,6}$ \And Padma Kaul$^{1,3}$ 
\And 
\\[-5pt]
$^1$ Canadian VIGOUR Centre, University of Alberta, Edmonton, Canada \\
$^2$ Department of Computing Science, University of Alberta, Edmonton, Canada \\
$^3$ Department of Medicine, University of Alberta, Edmonton, Canada\\
$^4$ Alberta Health Services, Alberta, Canada\\
$^5$ Department of Engineering, University of Alberta, Edmonton, Canada\\
$^6$ Alberta Machine Intelligence Institute, Edmonton, Canada
\And
weijie2@ualberta.ca \And kalmady@ualberta.ca \\
}

\begin{document}

\maketitle

\begin{abstract}
  Pandemic outbreaks such as COVID-19 occur unexpectedly, and need immediate action due to their potential devastating consequences on global health. Point-of-care routine assessments such as electrocardiogram (ECG), can be used to develop prediction models for identifying individuals at risk. However, there is often too little clinically-annotated medical data, especially in early phases of a pandemic, to develop accurate prediction models. In such situations, historical pre-pandemic health records can be utilized to estimate a preliminary model, which can then be fine-tuned based on limited available pandemic data. This study shows this approach -- pre-train deep learning models with pre-pandemic data -- can work effectively, by demonstrating substantial performance improvement over three different COVID-19 related diagnostic and prognostic prediction tasks. Similar transfer learning strategies can be useful for developing timely artificial intelligence solutions in future pandemic outbreaks.
\end{abstract}

\section{Introduction}

Coronavirus disease (COVID-19), the infectious disease caused by the SARS-CoV-2 virus, led to a worldwide pandemic beginning at end of 2019, with 546,363,985 confirmed cases and 6,336,802 deaths as of July 2022 \cite{world2022covid}. Initially regarded as a respiratory infection, COVID-19 is now known to affect all major systems in the body, including the cardiovascular system by causing myocardial damage, vascular inflammation, plaque instability, and myocardial infarction \cite{inciardi2020cardiac}. COVID-19 patients have shown ischemic changes, rate, rhythm abnormalities, and conduction defects in their \textit{electrocardiograms} (ECG) \cite{kaliyaperumal2022electrocardiographic}. In this context, machine-learned models that use ECG data could prove helpful in recognizing patients who require urgent definitive management. 

Moreover, the presence of cardiac involvement in COVID-19 is known to result in poor prognosis and adverse outcomes \cite{magadum2020cardiovascular}. Hence, monitoring the cardiac function is crucial for prompt identification of the appropriate actions. The ECG, which provides important information about the structure and electrical activity of the heart, is a simple point-of-care diagnostic tool \cite{maleki2014use} that can be employed to assess cardiovascular involvement in COVID-19 patients. Despite the massive impacts of the novel coronavirus, most existing studies employing machine learning methods to predict COVID-19 outcomes using ECGs are conducted in limited data sizes (COVID-19 diagnosis prediction using a public dataset of ~2000 scanned ECG images \cite{sobahi2022attention, bassiouni2022automated, gomes2022covid, attallah2022ecg, attallah2022intelligent, irmak2022covid, rahman2022cov} mortality prediction in COVID-19 patients with 882 \cite{van2022electrocardiogram} and 1148 \cite{sridhar2022identifying} ECG traces). In general, there are no prediction models that are trained and tested on population-scale ECG voltage time-series data to identify who is most likely to test positive for COVID-19; and among those who have COVID-19, who is most likely to suffer adverse consequences.

In this study, we use population-scale administrative health records and large ECG datasets, with two year coverage from the start of the pandemic (January 2020 to December 2021) to develop deep learning (DL) models to predict diagnostic and prognostic outcomes in COVID-19. Accordingly, we develop DL models to: (1) identify patients with COVID-19; and (2) identify COVID-19 patients who are at higher risk of mortality. Further, we demonstrate that transfer learning from pre-pandemic clinical records (Feb 2007 to Dec 2019) can be utilized to improve the prediction performance of COVID-19 models.

\section{Methods}

In this study, we explored three COVID-19 related prediction tasks using 12-lead ECG tracing, age and sex of patients, including:

\begin{enumerate}
    
    \item Diagnosis prediction with a binary classification framework (predicting if a given patient has COVID-19 or not).

    \item Mortality prediction with a binary classification for 30-day mortality (i.e. if a patient with COVID-19 will die within 30 days after the acquisition of ECG).
    
    \item Mortality prediction on a continuous time-scale using an individualized survival distribution (ISD) for each patient with COVID-19.
    
\end{enumerate}

The province of Alberta, Canada has a single-payer (Ministry of Health: Alberta Health) and single-provider (Alberta Health Services) healthcare system. The 4.4 million residents have universal access to hospital, ambulatory, laboratory, and physician services. For this study, ECG data were linked with the provincial-scale administrative health databases using unique patient health numbers. These databases included information on inpatient hospitalizations, outpatient clinic and emergency department (ED) visits, demographic information, and date of death. We used standard ECG voltage-time series, sampled at 500 Hz for 10 seconds for each of 12 leads (500 x 10 x 12 voltage measurements per ECG). We also used age and sex of the patients as additional features. An ECG record was linked to a healthcare episode if the acquisition date was within the timeframe between the admission date and discharge date of an episode. The COVID-19 diagnosis was identified using the ICD-10 codes (U071, U073) \cite{wu2022validity} in the primary or secondary diagnosis fields of a healthcare episode linked to a particular ECG. This study received ethics approval from the Health Research Ethics Board at the University of Alberta (Pro00120852).

\begin{figure}[htp!]
% \advance \leftskip-3.7cm
\centering
\includegraphics[width=\textwidth]{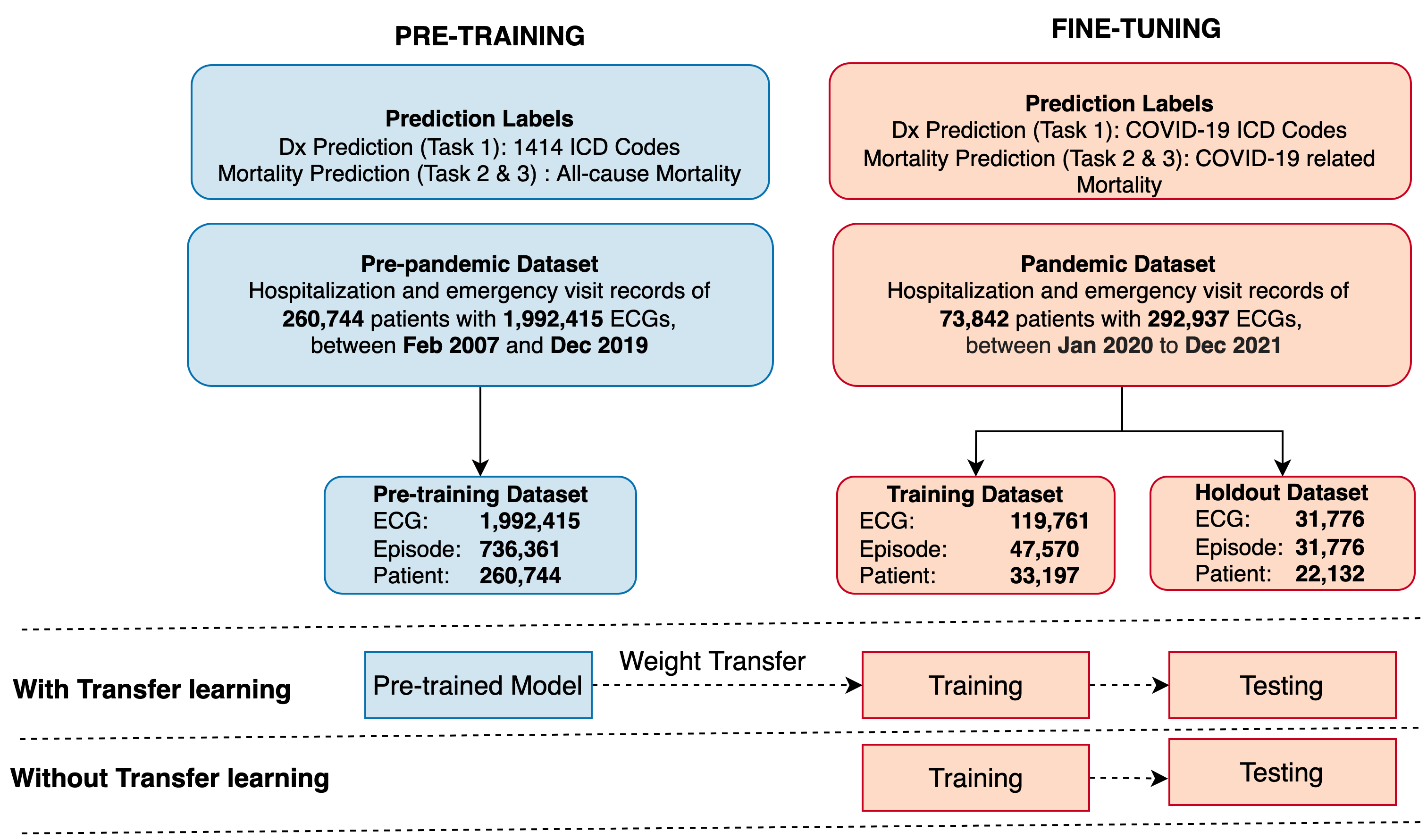}
\caption{Datasets used for pre-training and then fine-tuning the COVID-19 prediction tasks}
\label{fig:flowchart}
\label{table1}
\end{figure}

Our dataset consisted of 73,842 patients during the COVID-19 pandemic period (Jan 2020 - Dec 2021) who had taken 292,937 ECG tests overall. We excluded ECGs that could not be linked to healthcare episodes, patients < 18 years of age, and ones with poor signal quality. Then, we split our ECG dataset into the development dataset (random 60\%: 33,197 patients with 119,761 ECGs, used for training and internal validation) and holdout dataset (remaining 40\%: 22,132 patients with 80,097 ECGs), while ensuring that ECGs from the same patient were not shared between the sets. In addition, we also used a retrospective pre-pandemic dataset that included 260,774 patients with 1,992,415 ECG obtained between Feb 2007 to Dec 2019; see Figure \ref{fig:flowchart}).

First, we trained our models within the pandemic time frame (Jan 2020 to Dec 2021) and then we extended our training to pre-pandemic time frame (Feb 2007 to Dec 2019) to compare the performance improvement. Since the data on COVID-19 diagnosis was not available before the pandemic, we used a model for predicting over 1414 ICD-10 codes (multi-label classification model previously developed on pre-pandemic data evaluated using area under the receiver operating characteristic curve (AUROC) \cite{sun2021ecg}) as a pre-training model for the COVID-19 diagnosis task. Similarly, since data on COVID-19 related deaths were not available before the pandemic, we used the all-cause mortality prediction model (again, previously developed on pre-pandemic data,  unpublished) as a pre-training model for the COVID-19 mortality task. Both of these pre-trained models were then fine-tuned on COVID-19 specific predictions using the development set (60\% of pandemic data) and evaluated on hold-out set (40\% of pandemic data); see Figure \ref{fig:flowchart}.

For the classification tasks, we implemented a convolutional neural network (CNN) based on the ResNet, consisting of a convolutional layer, 4 residual blocks with 2 convolutional layers per block, followed by a dense layer (total of 6,400,433 model parameters). We used batch normalization \cite{ioffe2015batch}, ReLU and dropout \cite{hinton2012improving} after each convolutional layer. Our architecture was based on a previously published large-scale study to identify abnormalities in 12-lead ECGs \cite{ribeiro2020automatic} with some modifications to accommodate tabular data input and binary output. Each ECG instance was loaded as a 12 x 4096 numeric matrix. Sex (binary) and age features (real valued) were passed to a 10-hidden-unit layer, then concatenated with the dense layer, and finally passed to a sigmoid function to produce the output (Task 1: Probability of diagnosis; or Task 2: Probability of 30-days mortality). Binary cross-entropy was used as the loss function with the initial learning rate of $10^{-3}$, Adam optimizer \cite{kingma2014adam}, ReLU activation function, kernel size of 16, batch size of 512, and dropout rate of 0.2 with other hyperparameters set to default. The learning rate was reduced to $10^-6$ if there was no improvement in tuning loss for nine consecutive epochs, and the learning process was stopped if the loss did not reduce for nine epochs. For individual survival prediction tasks (Task 3), we designed an end-to-end ISD algorithm, which is consistent with the ResNet structure (the sigmoid was replaced with ReLu function) followed by 3 fully connected layers and a Multi Task Logistic Regression \cite{yu2011learning} (MTLR, see appendix for overview) module.

In order to implement transfer learning, (a) in the classification algorithm, we froze all previous layers' weights and added one fully connected layer and; (b) in the ISD algorithm, we froze all layers except the MTLR layers, during the training process. The models were implemented using PyTorch 1.11 in Python 3.8. We trained all our models with 8 Tesla V100-SXM2 GPUs and 32 GB of RAM per GPU.

Our goal is to ultimately generate a prediction system that could be employed at the point of care – i.e when the patient’s first ECG is acquired during an ED visit or hospitalization. We therefore used the first ECG of each episode in the holdout set to evaluate our diagnosis model, and first ECG of a random COVID-19 episode for each patient in the holdout set to evaluate our mortality models. The validation performance of these models were compared using AUROC or the Concordance-index (C-index) and bootstrapped 95\% confidence intervals with 1000 replicates (CI). Additionally, we report Area Under the Precision-recall Curve (AUPRC), and Average Precision (AP), F1, specificity, recall, precision, accuracy, and Brier score for binary classification as well as L1 marginal loss, and L1 hinge loss for ISD prediction.

\begin{figure}[ht]
\centering
\includegraphics[width=135mm]{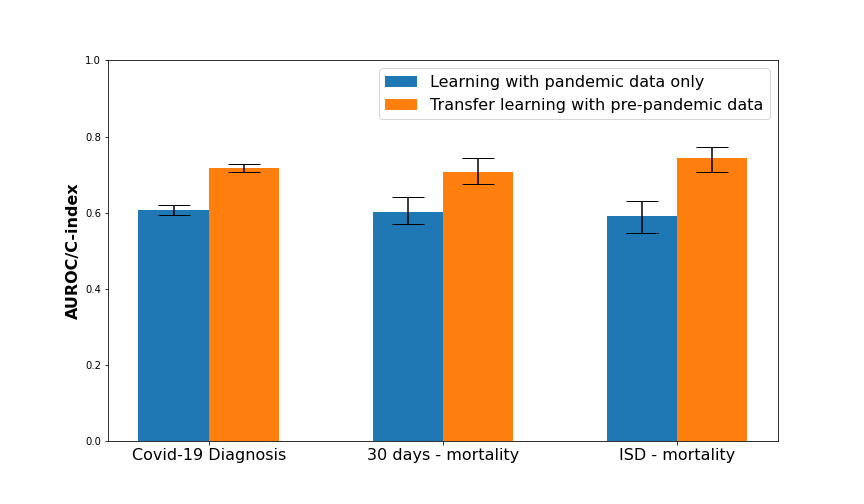}
\caption{}{Comparison of AUROC / C-Index for COVID-19 prediction models with and without pre-pandemic transfer learning. Note error bars are 95\%CI.\protect\footnotemark}
\label{fig:Group_AUROC}
\end{figure}

\footnotetext{AUROC: area under the receiver operating characteristic curve; C-Index: concordance index; CI: Confidence Interval}

\section{Results}

Supplementary Table \ref{table:performance_table} and Figure \ref{fig:Group_AUROC} show the evaluation performance of three models, with and without pre-pandemic transfer learning. After applying the exclusion criteria, out of the total of 199,858 ECGs acquired from 55,329 patients in the pandemic period, 3,452 (6.24\%) were tested positive for COVID-19. The diagnosis classification model trained solely with a development dataset from this period showed AUROC of 60.7\%. When the same algorithm was pre-trained with ICD-wide diagnosis from the pre-pandemic time frame, the COVID-19 diagnosis performance improved significantly by 11.1\% to 71.8\%.

Further, 407 (11.79\%) patients out of the total 3,452 patients who were diagnosed with COVID-19 died within 30 days (See Kaplan Meier curve, Figure \ref{fig:KM_curve}). Similar to the diagnosis task, performance of our 30-days mortality prediction improved significantly by 10.6\% from AUROC of 60.1\% to 70.7\% when the model was pre-trained with pre-pandemic all-cause 30-day mortality.

\begin{figure}[ht]
\centering
\includegraphics[width=135mm]{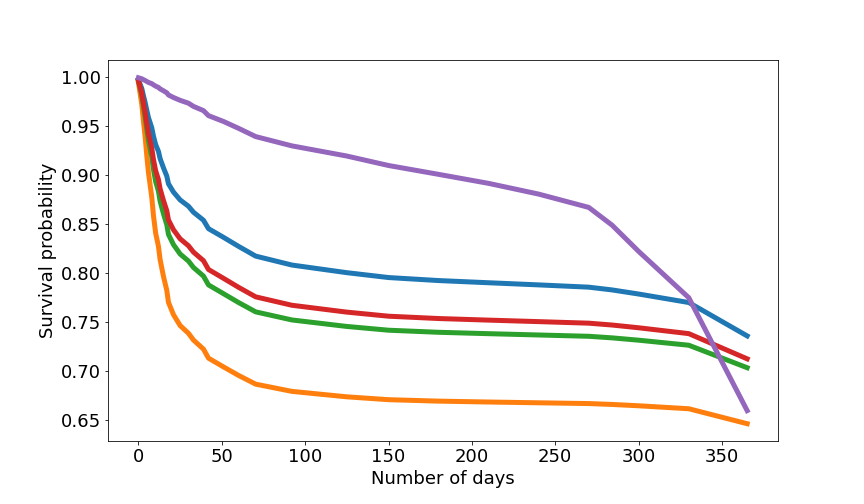}
\caption{Individual survival curve predictions for 5 sample COVID-19 patients}
\label{fig:ISD_representation}
\end{figure}

Predicted survival curves using our ISD model for representative patients with COVID-19 are shown in Figure \ref{fig:ISD_representation}. We observed the similar trend with ISD prediction as well, with the significant C-index  improvement of 15.3\% (from 59\% to 74.3\%) using transfer learning.

% \begin{figure}[htp!]
% \centering
% \includegraphics[width=152mm]{images/survival_representative.png}
% \caption{Kaplan Meier Curve for COVID-19 patients in the study cohort}
% \label{fig:ISD_representation}
% \end{figure}

\section{Discussion and Conclusion} 

The generalization capability of the convolutional DL models strongly depend on the training data size and the architecture complexity \cite{yamashita2018convolutional}. As seen in this study, training a DL model using smaller training data with randomized weights initialization, can have high variance and low performance on a test set. Transfer learning is often used to learn a model for a “target” domain when we have a limited number of training instances for that domain, but have many training instances for a related “source domain”. Here, we first pre-train a model on the source dataset, then fine-tune that learned model on the target dataset. This idea of transfer learning which was initially developed in DL-based computer vision applications \cite{han2021pre}, is now widely applied in various medical domains \cite{kermanyidentifying, karimi2021transfer}. Earlier this year, Leur et al \cite{van2022electrocardiogram} showed that transfer learning of a deep neural network to raw ECGs from a very heterogeneous source performed similarly to the time-consuming human interpretation of the ECG for mortality prediction in COVID-19 patients. This motivated us to use the transfer learning method of first pre-training our DL models on a temporally disjoint dataset of ECGs acquired before the pandemic out-break, and then subsequently continued to train them on our target task and target dataset after the pandemic outbreak. This often produces a better representation of ECG features in lower layers of DL, which then would yield better initialization of model parameters for training on the target task \cite{yosinski2014transferable}. For the target task of COVID-19 diagnosis, we chose 1414 different ICD-10 diagnosis codes as labels for our source task to capture global changes in ECG signals that are associated with various medical conditions in general. Also, we expect that our pre-train models have built a relevant internal representation of the patients, which can then be fruitful even if the outcome changes.

Recall or sensitivity is one of the important metrics for COVID-19 diagnosis, since a lot of false negatives would result in many undiagnosed patients who can potentially spread the virus. COVID-19 diagnosis task with transfer learning showed 11.1\% improvement in AUROC, suggesting that transfer learning can result in a model that produces prediction probabilities which are more concordant with the observed COVID-19 status. Interestingly, the resultant model had 22\% improvement in specificity, but 9.8\% decrease in recall. In this context, it is noteworthy that we have not considered the differential costs of misclassification associated with false positives and false negatives in these early stages of model development. Other important factors are the prevalence of the condition in the target population as well as costs associated with running ECG tests and prediction algorithms at the point of care. These considerations should be taken into account to estimate the optimal cut-point for binarizing the prediction probabilities for COVID-19 diagnosis during the deployment phase \cite{drummond2006cost}.

There is limited reliable medical data available during pandemic outbreaks, especially in the early phases. This poses significant challenges for DL applications targeted at identifying individuals at risk or aiding clinical policy decisions at population level. Historical electronic health records can be leveraged to estimate our best guess in such situations. This study demonstrates an application of pre-pandemic transfer learning by showing substantial improvement over 3 different COVID-19 related diagnostic and prognostic prediction tasks. We expect that similar modeling strategies can be useful in developing timely learning tools during pandemic outbreaks in the future. However, despite our models being validated on large ECG datasets, the wide scale generalizability of such transfer learning approaches in pandemic situations warrants further investigation. 

\bibliographystyle{unsrt}
\bibliography{references}

\newpage
\begin{center}
  \Large\textbf{Appendix}\\
\end{center}
\appendix

% \section{Figure}
% \counterwithin{figure}{section}

\section{Figure}
\counterwithin{figure}{section}

\begin{figure}[htbp]
\centering
\includegraphics[width=152mm]{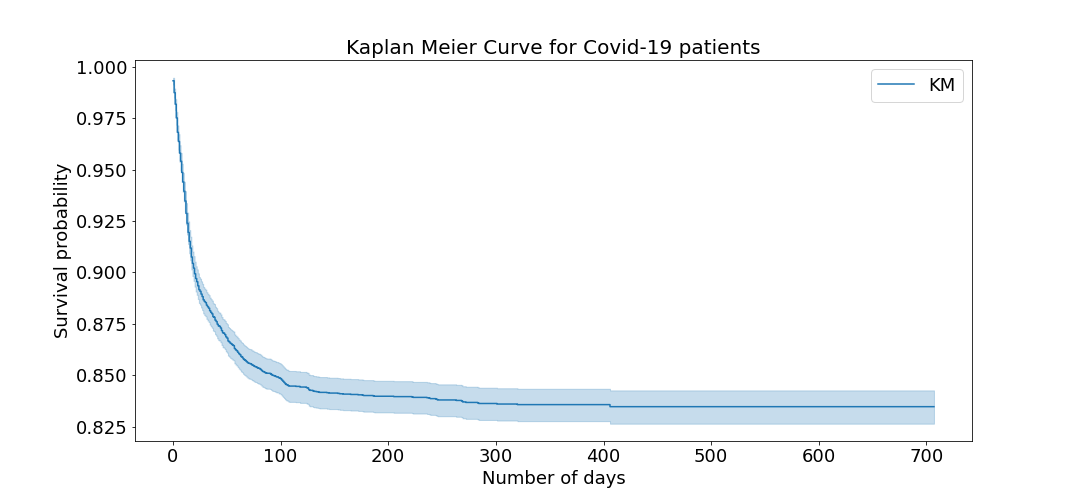}
\caption{Kaplan Meier Curve for patients with COVID-19 in the study cohort}
\label{fig:KM_curve}
\end{figure}

\section{Overview of MTLR}

Consider modeling the probability of survival of patients at each of a vector of time points $\tau = [t_1, t_2, \ldots, t_m]$. In our study $\tau$ is the union of $\sqrt{N_{uncensored}}$ and month-wise in 1 years. We can set up a series of logistic regression models: For each patient, represented as $\vec{x}$, 

\begin{equation}
% \begin{aligned}
P(T \geq t_i \,|\, \vec{x}) =  \left(1 + \exp(\vec{\theta_{i}}\cdot \vec{x} )\right)^{-1}
% \end{aligned}
\end{equation}

where $\vec{\theta_i}$ are the time-specific feature vectors. While the input features $\vec{x}$ stay the same for all these classification tasks, the binary labels $y_i = [T\geq t_i]$ can change depending on the threshold $t_i$.\%
We encode the survival time $d$ of a patient as a binary sequence $y = y(d) = (y_1, y_2, \ldots, y_m)$, where $y_i = y_i(d) \in \{0,1\}$ denotes the survival status of the patient at time $t_i$, so that $y_i = 0$ (no death or readmission event yet) for all $i$ with $t_{i} < d$, and $(y_i = 1)$ (death) for all $i$ with $t_i \geq d$.
\\\\
Here there are $m+1$ possible legal sequences of the form $(0,0,\ldots,1,1,\ldots,1)$, including the sequence of all 0’s and the sequence of all 1’s. The probability of observing the survival status sequence $y = (y_1, y_2, \ldots, y_m)$ can be represented as: 
\begin{equation}
% \begin{aligned}
P_\Theta(Y\!\! =\!\! (y_1, y_2, \ldots, y_m) \,\,|\,\, \vec{x}) = \frac{\exp(\sum_{i=1}^m y_i \times \theta_{i}\cdot\vec{x} )}
     {\sum_{k=0}^m \exp(f_{\Theta}(\vec{x}, k))}, 
% \end{aligned}
\end{equation} 
where $\Theta = (\theta_{1}, \ldots, \theta_{m})$, and $f_{\Theta}(\vec{x}, k) = \sum_{i=k+1}^m (\theta_{i}\cdot\vec{x})$ for $0\leq k\leq m$ is the score of the sequence with the event occurring in the interval $[t_k, t_{k+1})$ before taking the logistic transform, with the boundary case $f_{\Theta}(\vec{x}, k)= 0$ being the score for the sequence of all `0’s. Given a dataset of $n$ patients $\{ \vec{x_r}\}$ with associated time of events $\{ d_r \}$, we find the optimal parameters (for the MTLR model) $(\Theta^*)$ as 
\begin{equation}
% \begin{aligned}
\Theta^*\ =\
\arg\max_{\Theta} 
\sum_{r=1}^n \left[\sum_{i=1}^m y_j(d_r)(\theta{i}\!\cdot\!\vec{x_r})\! - \log \sum_{k=0}^m \exp f_{\Theta}(\vec{x_r},k) \right]
  -  
\frac{C}{2}\!\sum_{j=1}^m\! \|\theta_{j}\|^2\! 
% +\! \frac{C_2}{2}\sum_{j=1}^{m-1}\! \|\theta_{j+1}\!-\!\theta_j\|^2\!
% \end{aligned}
\end{equation}
where the $C$ (for the regularizer) is found by an internal cross-validation process. 
There are many details here –  to insure that the survival function starts at 1.0, and decreases monotonically and smoothly until reaching 0.0 for the final time point; to deal appropriately with censored patients; to decide how many time points to consider ($m$); and to minimize the risk of overfitting (by regularizing), and by selecting the relevant features. 
\\\\
The paper by Yu et al. provides the details. C.-N. Yu, R. Greiner, H.-C. Lin, and V. Baracos. Learning patient-specific cancer survival distributions as a sequence of dependent regressors. In NIPS, 2011\cite{yu2011learning}.

% \newpage
\section{Performance table}
\counterwithin{table}{section}

\begin{table}[htbp]
% \advance\leftskip-1.5cm
\begin{tabular}{|l|ll|ll|l|ll|}
\hline
\rule{0pt}{10pt}Task $\rightarrow$                                                       & \multicolumn{2}{l|}{COVID-19 Diagnosis} & \multicolumn{2}{l|}{30 days (binary mortality)} & Task $\rightarrow$                                                        & \multicolumn{2}{l|}{ISD mortality}         \\ \hline
\rule{0pt}{20pt}\begin{tabular}[c]{@{}l@{}}Transfer \\ learning\\ $\rightarrow$\end{tabular} & \multicolumn{1}{l|}{No}       & Yes     & \multicolumn{1}{l|}{No}           & Yes         & \begin{tabular}[c]{@{}l@{}}Transfer \\ learning\\ $\rightarrow$ \end{tabular} & \multicolumn{1}{l|}{No}        & Yes       \\ \hline
\rule{0pt}{10pt}AUROC                                                                                    & \multicolumn{1}{l|}{60.70\%}  & 71.80\% & \multicolumn{1}{l|}{60.10\%}      & 70.70\%     & C index                                                                                  & \multicolumn{1}{l|}{59.00\%}   & 74.30\%   \\ \hline
\rule{0pt}{10pt}AUPRC                                                                                    & \multicolumn{1}{l|}{6.49\%}   & 13.00\% & \multicolumn{1}{l|}{15.74\%}      & 24.83\%     & L1 marginal loss                                                                         & \multicolumn{1}{l|}{2047.9808} & 2057.6506 \\ \hline
\rule{0pt}{10pt}AP                                                                                       & \multicolumn{1}{l|}{6.54\%}   & 13.10\% & \multicolumn{1}{l|}{16.03\%}      & 25.11\%     & L1 hinge loss                                                                            & \multicolumn{1}{l|}{79.2981}   & 74.9171   \\ \hline
\rule{0pt}{10pt}F1                                                                                       & \multicolumn{1}{l|}{10.70\%}  & 16.40\% & \multicolumn{1}{l|}{24.84\%}      & 30.28\%     &                                                                                          & \multicolumn{1}{l|}{}          &           \\ \hline
\rule{0pt}{10pt}specificity                                                                              & \multicolumn{1}{l|}{57.70\%}  & 79.70\% & \multicolumn{1}{l|}{77.25\%}      & 76.41\%     &                                                                                          & \multicolumn{1}{l|}{}          &           \\ \hline
\rule{0pt}{10pt}recall                                                                                   & \multicolumn{1}{l|}{58.50\%}  & 48.70\% & \multicolumn{1}{l|}{40.69\%}      & 52.41\%     &                                                                                          & \multicolumn{1}{l|}{}          &           \\ \hline
\rule{0pt}{10pt}precision                                                                                & \multicolumn{1}{l|}{5.90\%}   & 9.83\%  & \multicolumn{1}{l|}{17.88\%}      & 21.29\%     &                                                                                          & \multicolumn{1}{l|}{}          &           \\ \hline
\rule{0pt}{10pt}accuracy                                                                                 & \multicolumn{1}{l|}{57.70\%}  & 78.40\% & \multicolumn{1}{l|}{73.27\%}      & 73.80\%     &                                                                                          & \multicolumn{1}{l|}{}          &           \\ \hline
\rule{0pt}{10pt}Brier score                                                                              & \multicolumn{1}{l|}{4.16\%}   & 4.15\%  & \multicolumn{1}{l|}{9.72\%}       & 9.53\%      &                                                                                          & \multicolumn{1}{l|}{}          &           \\ \hline
\rule{0pt}{10pt}P+ (test)                                                                                & \multicolumn{1}{l|}{1380}     & 1380    & \multicolumn{1}{l|}{145}          & 145         &                                                                                          & \multicolumn{1}{l|}{}          &           \\ \hline
\rule{0pt}{10pt}P- (test)                                                                                & \multicolumn{1}{l|}{30377}    & 30377   & \multicolumn{1}{l|}{1191}         & 1191        &                                                                                          & \multicolumn{1}{l|}{}          &           \\ \hline
\rule{0pt}{10pt}P+ (all)                                                                                 & \multicolumn{1}{l|}{9090}     & 9090    & \multicolumn{1}{l|}{1064}         & 1064        &                                                                                          & \multicolumn{1}{l|}{}          &           \\ \hline
\rule{0pt}{10pt}P- (all)                                                                                 & \multicolumn{1}{l|}{190768}   & 190768  & \multicolumn{1}{l|}{7836}         & 7836        &                                                                                          & \multicolumn{1}{l|}{}          &           \\ \hline
\end{tabular}
\caption{ }{Performance of COVID-19 prediction models with and without pre-pandemic transfer learning. p(+) refers to the number of positive instances for test (p+ (test)) and full (train + test) dataset (p+ (all)).\protect\footnotemark}
\label{table:performance_table}
\end{table}

\footnotetext{AUROC: area under the receiver operating characteristic curve; AUPRC: Area Under the Precision-recall Curve; AP: Average Precision; P+/- (test): positive/negative instances in test set; P+/- (all): positive/negative instances in all sets; ISD: individual survival distribution; C index: concordance index}

\end{document}